\documentstyle[preprint,aps]{revtex}
\preprint {IMSc-94/42}
\begin{document}
\draft
\title{
Haldane exclusion statistics and the charge fractionalisation in
chiral bags
}

\author{ R.K.Bhaduri$^{a,b}$ , M. V. N. Murthy$^a$
 and R. Shankar$^a$}

\address
{$^a$The Institute of Mathematical Sciences, Madras 600 113, India.\\
$^b$ Department of Physics and Astronomy, McMaster University, Ont.
L8S 4M1, Canada.\\}
\date{\today}
\maketitle

\begin{abstract}
It is proposed that the phenomenon of charge fractionalisation of the
spatially confined particle in a topological chiral bag may
be interpreted as a manifestation of the exclusion statistics proposed
by Haldane.  The fractional exclusion statistics parameter is just the
fractional baryon charge $Q$ of the particle in this case. We also
state the necessary conditions for Haldane fractional to occur for
systems in higher dimensions.
\end{abstract}


\narrowtext
Haldane\cite{haldane} has recently proposed a new exclusion statistics
for quasi particles that has attracted a lot of
attention\cite{ouvry,canjohn,ms,wu}. His definition holds for systems
with finite dimensional Hilbert spaces and is motivated by  physical
examples, such as quasi-particles in the fractional quantum Hall
systems and spinons in antiferromagnetic spin chains.  Haldane first
defines $d_N$, the dimension of the single particle space when there
are $N$ particles present, as follows:  Consider the $N$ particle wave
function $\psi(x_1,x_2,...,x_N)$.  Fix any $N-1$ of the variables at
any $N-1$ points and look upon the wavefunctions as  functions of the
remaining single particle variable.  The dimension of the space
spanned by these
functions, $d_N$, then defines the single particle Hilbert space of
dimension $d_N$ in the
presence of $N-1$ other partices. To clarify the definition, consider
particles on a lattice of dimension $d_1$.  For fermions, if there are
N particles and  we fix the position of $N-1$ particles on the
lattice, then the remaining single particle can only occupy any of the
$d_1-(N-1)$ remaining positions due to Pauli blocking. Thus the
fermionic single
particle dimension as defined by Haldane is $d_N^F = d_1 -(N-1)$.
Bosons on the other hand can occupy any of the $d_1$ sites
irrespective of how many other particles are there.  Therefore
$d_N^B=d_1$.  Haldane then considers systems with a generalized Pauli
blocking whose single particle space is given by,
\begin{equation}
d_N = d_1 - g(N-1),
\end{equation}
where $d_N$, as stated above, is the number of states accessible to a
particle in the
presence of $N-1$ other particles.The exclusion statistics
parameter  $g$ is then given   by
\begin{equation}
g = -\frac{\Delta d}{\Delta N}
\end{equation}
where $\Delta d$ is the change in the dimension of the single particle
space and $\Delta N$ is the change in the number of particles when the
size of the system and the boundary conditions are unchanged. Thus
$g$  is a measure of the (partial) Pauli blocking in the system,
$g=0(1)$ corresponds to bosons (fermions).  Recently this new
definition of statistics has been generalised to the case of systems
with infinite dimensional Hilbert spaces\cite{ms}.  It was found that
in such cases the exclusion statistics of a system of interacting
fermions(or bosons) was non-trivial if the interaction was such that
the addition of a particle to the system caused a scale invariant
shift to the energy all the of other particles in the system.  Interacting
systems in one-dimension, like the quasiparticles of
Calogero-Sutherland Model, are generic examples of this phenomenon.
Anyons (as defined by the exchange phases) in two dimensions is
another example\cite{ms}.     In this letter, by considering a class of
relativistic models we are able to relate the Haldane statistical
interpolating factor $g$ to the fractional charge. We do this by first
demonstrating the relation between fractional baryon charge and the
Haldane statistical parameter in a  1+1 dimensional chiral bag model
due to Zahed\cite{zahed} and then generalize the result to the chiral
bag model in 3+1 dimensions. This connection between the statistical
parameter and the fractional charge indicates that $g$ may have
topological significance.   A hint of this already emerges from the
elegant proof\cite{comtet} that the second  virial coefficient   of a
nonrelativistic anyon gas in two space dimensions  is determined by
the axial anomaly of a 1+1- dimensional fermionic system gauged by a
vortex vector field.  This is so since $g$ itself has been
shown\cite{ms} to be determined by the high temperature limit of the
second virial coefficient in cases where a virial expansion exists.
In the end we also remark about an analogous phenomenon in the Kondo
system.

To make this connection between the Haldane parameter $g$ and
fractional charge we first give a regulated definition of $d_N$
applicable to  systems of interacting fermions where the many particle
spectrum admits an interpretation in terms of effective single
particle levels (which may be $N$ dependent).  Suppose these single
particle energies are denoted by $\epsilon_n^N$, then the regulated
definition of
$d_N(\beta)$ is given by,
\begin{equation}
d_N(\beta) = \sum_{n(unocc)} e^{-\beta\epsilon_n^N},
\end{equation}
where $\beta$ is the inverse temperature and $d_N = \lim_{\beta
\rightarrow 0} d_N(\beta)$.  The sum is taken over all the unoccupied
levels.The occupancies could correspond to any  state in the
$N-$particle sector.  For convenience, we will always work with the
ground state.  We can then compute the quantity,
\begin{equation}
\Delta d = \lim_{\beta\rightarrow 0}
(d_{N+1}(\beta)-d_N(\beta)).\label{regdef}
\end{equation}
If this limit exists and is $N$ independent, then the system can be
interpreted in terms of Eq.(1) and we have $g=-\Delta d$.

To be more specific we first apply this definition to the
Calogero-Sutherland Model(CSM)\cite{csm}  which can be looked upon
either as a system of interacting fermions or ideal exclusion anyons
in one dimension\cite{ms1,wu1}.  It is a system of non-relativistic
fermions in one dimensions confined in a harmonic oscillator potential
with frequency $\omega$ and interacting via an inverse-square
potential $\sum_{i<j} \lambda (\lambda+1)/(r_{ij})^2$ , where $i,j$
runs over all particle positions and $\lambda $ is the interaction
parameter. Note that the range of $\lambda$ is limited to $ \lambda
\ge -1/2 $ in the fermionic basis. Otherwise the system does not lead
to bound states. The states can be labelled by sets of occupied
harmonic oscillator levels.  The $N$ particle energy  spectrum is
given by,
\begin{equation}
E[\{n_i\}] = E_0[\{n_i\}] + \lambda \omega
\frac{N(N-1)}{2},
\end{equation}
where $E_0$ is the energy of the N non-interacting fermionic system.
This can be interpreted in terms of effective single particle levels
$\epsilon_n^N =\omega (n +\frac{1}{2}) + \lambda \omega (N-1)$. (Note
that the interaction
energy is divided by a factor 2 in the total energy to avoid
overcounting.) The $d_N(\beta)$ can now be calculated using eq.(3) and
we find that,
$$ d_N(\beta) =
\frac{e^{-\beta\omega(\frac{1}{2}+(\lambda+1)(N-1))}}{1-e^{-\beta\omega}}.
$$
Using eq.(4) it follows, in the high temperature limit $(\beta
\rightarrow 0)$, that $\Delta d = -(1+\lambda)$ and hence we obtain
$g=1+\lambda$ which is known to be the correct result and demonstrates
that our approach is consistent.

As advertised, we now relate the Haldane statistical parameter $g$ to
the vacuum charge(under certain conditions) in a relativistic system.
The model with which we first illustrate this connection is  the
(1+1)- dimensional chiral bag model due to Zahed\cite{zahed}.  This
simple model is chosen because it is completely solvable, and brings
about the main features of the (3+1)-dimensional chiral bag in this
context.  For our  purposes, it is sufficient to consider a
flavour-singlet Dirac particle with unit baryon number or charge, N=1.
The model is characterised by dividing the spatial dimension into
three sections, the exterior solitonic coat, the interior bag with the
confined fermion and the boundary between the two where the fermion
interacts with the soliton.   In the notation of ref.\cite{zahed}, the
Lagrangian density is given by,
\begin{equation}
{\cal L} = [\bar q i \frac{1}{2}\gamma_{\mu}\partial^{\mu}q -B]\theta_v
- \frac{1}{2} \bar q e^{-i\theta \gamma_5} q \Delta_s + [\frac{1}{2}
(\partial _{\mu}\theta)(\partial^{\mu}\theta)
-k^2(1+\cos(\theta))](1-\theta_v), \label{lagrangian}
\end{equation}
where $\theta_v$ is a theta function at the boundary defined at a
radius R which ensures that the quarks remain inside the bag and
$\Delta_s$ is the surface delta-function.

The one baryon number solution then consists of the $\theta$ field
being the 1-kink solitonic solution of the mesonic action and with one
extra fermion in the bag.  As emphasised by Zahed, the chiral boundary
condition gives a global rotation to the solutions of the Dirac
equation and causes all the energy levels to shift by $\theta (R)/2R$.
They are given by,
\begin{equation}
\epsilon_n^1 = \frac{(2n+1)\pi}{4R} +\frac{ \theta(R)}{2R} =
\frac{\pi}{2R} (n+ {1\over 2} +\frac{\theta(R)}{\pi}).
\end{equation}
For the purposes of regularization we also define the so called
``empty bag''  as follows: Consider the situation in which particles
are confined in a region of radius R. The energy levels of the
particles are obtained by solving the Dirac equation and are given by,
\begin{equation}
\epsilon_n^0 = \frac{(2n+1)\pi}{4R}.
\end{equation}
Now consider a situation where all the negative energy states are
fully occupied with no valence particle present. We refer to this
system as the ``empty bag''. The states inside the empty bag therefore
correspond to the solutions of the Zahed model (eq.(6)) in the zero
baryon number sector.  The solution of the sine-Gordon equation in
this sector is $\theta(R)=$ constant which may be set equal to zero.
Hence there is no chiral coupling.
Thus for a fixed R, we have an $N-$particle($N \rightarrow \infty$)
system which is nothing but the filled negative energy sea (the zero
baryon number sector). As soon as the valence particle is introduced,
the solitonic field deforms due to the chiral coupling at the boundary
R and we regard this as the $(N+1)$-particle system (one baryon number
sector). We are now interrested in computing $\Delta d$ a la eq.(4).

The total number of states of the empty bag, for a fixed boundary at
R, is obtained as  the sum over occupied and unoccupied levels and
is given by,
\begin{equation}
d_{+}(\beta) + d_{-}(\beta) = \sum_{n=-\infty}^{n=\infty} e^{-\beta
\frac{\pi}{ 2R}|n+\frac{1}{2}|} = \frac{4R}{\pi\beta} + O(\beta),
\end{equation}
where - and + subscripts denote the occupied and unoccupied states.
We now consider placing a valence particle keeping R the same as
before. The particles are now free in the finite confined region  but
interact with the meson fields outside at the boundary. The total
number of states is given by,
\begin{equation}
d^v_{+}(\beta) + d^v_{-}(\beta) = \sum_{n=-\infty}^{n=\infty} e^{-\beta
\frac{\pi}{ 2R}|n+\frac{1}{2}+\frac{\theta(R)}{\pi}|} =
\frac{4R}{\pi\beta} + O(\beta),
\end{equation}
where the dependence on the chiral angle occurs at
$O(\beta)$. Therefore in the high temperature limit,
\begin{equation}
\lim_{\beta \rightarrow 0} d_{+}(\beta) + d_{-}(\beta) =
\lim_{\beta \rightarrow 0} d^v_{+}(\beta) + d^v_{-}(\beta)
\end{equation}
Hence the change in the number of available single particle states
with the addition of a valence particle to the system is given by,
\begin{equation}
\Delta d =
\lim_{\beta \rightarrow 0} (d^v_{+}(\beta) - d_{+}(\beta)) =
-\lim_{\beta \rightarrow 0} (d^v_{-}(\beta) - d_{-}(\beta))=
\frac{1}{2}\lim_{\beta \rightarrow 0} (d^v_{+}(\beta) -
d^v_{-}(\beta)) =  -\frac{\theta(R)}{\pi}.
\end{equation}
Note that the last part of the above equation is just the spectral
asymmetry in the presence of the chiral coupling and $\theta(R)/\pi$
is precisely the result one obtains for the baryon charge
residing in the bag (see below). Therefore by definition (see eq.(1)
and eq.(4))
\begin{equation}
g = \frac{\theta(R)}{\pi}.
\end{equation}
Essentially what is happening in this model is that the
addition of one fermion in the bag causes the $\theta$- field to deform
to a soliton thus causing all the single particle energies to shift.

The above results for the Haldane statistical parameter may also be
interpreted in terms of the vacuum charge. In a relativistic problem,
the vacuum charge (or  equivalently the topological charge) of the
vacuum is given by the spectral asymmetry \cite{vacch},
\begin{equation}
N_v = -\frac{1}{2}\lim_{\beta \rightarrow 0} [\sum_{\epsilon_n \ge 0}
e^{-\beta |\epsilon_n|} - \sum_{\epsilon_n <0} e^{-\beta
|\epsilon_n|}].   \label{NV}
\end{equation}
For $\theta < \pi/2$, the vacuum charge is given by $ N_v =
\theta/\pi$ which is therefore the same as the Haldane $g$ parameter.
However, for $\theta > \pi/2$, the vacuum charge is given by
$N_v = \theta/\pi -1$.  The discontinuity at $\pi/2$ is caused by the
null mode at $\theta=\pi/2$.  However, in the computation of the net
baryon number  $Q$,  there is no discontinuity since the number of
particles in the vacuum is decreased by 1  for $\theta > \pi/2$ as the
valence particle has emerged out of the vacuum.  The excess charge is
then given by
\begin{eqnarray*}
Q = N_v = \theta/\pi, ~~~~~~ \theta < \pi/2 \\
Q = N_v +1 =\theta/\pi, ~~~\theta  \ge \pi/2.
\end{eqnarray*}
{\it Therefore the Haldane statistical parameter $g$ is the same as
the fractional baryon charge residing in the bag}. Note that the net
baryon charge is still unity since the sum of the baryon charge
carried by the chiral field outside and that inside the bag is equal
to 1.

We now consider the physically relevent 3+1 dimensional SU(2) chiral
bag model\cite{cbag}.  This model is exactly analogous to the Zahed
model described before.  As in that case the space is divided into
three regions.  Inside a sphere of radius R, we have non-interacting
quarks obeying the massless Dirac equation.  Outside the bag are
mesons described by the SU(2) Skyrme model.  The boundary conditions
satisfied by the Dirac fermions are
\begin{equation}
-i \vec \gamma. \hat n(\vec x) q(\vec x) = (U(\vec x)P_L + U^{\dag}
(\vec x) P_R) q(\vec x),
\end{equation}
where $q$ are the quark fields, $\vec x$ is any point on the boundary
such that $|\vec x| =R$, $U(\vec x)$ is the SU(2) valued field of the
Skyrme model and $P_L, P_R$ are left and right projection operators.
The one baryon state is described by putting one extra quark (per
colour) in the bag and a winding number 1 skyrmion outside the bag.
The skyrmion is given by the hedgehog solution
$ U(\vec x) = e^{i \theta(r) \vec x.\vec\tau /2}$
with $\theta(\infty) = 0,~~ \theta(0)=-\pi$. Goldstone and
Jaffe\cite{jaffe} have calculated the vacuum charge, $N_v$ of the
single particle Dirac hamiltonian with the boundary condition in
eq.(15). They find
\begin{eqnarray}
N_v(\theta)& =& \frac{1}{\pi}(\theta -sin\theta~ cos\theta), ~~-\pi/2 <
\theta < \pi/2\\
N_v(\theta +\pi)& =& N_v(\theta).
\end{eqnarray}
Hence, just as in the Zahed model, the excess charge is given by,
\begin{eqnarray}
Q &=& 1+N_v(\theta),~~~~ -\pi/2 \le \theta \le 0 \\
Q &=& N_v(\theta), ~~~~~~-\pi \le \theta \le  -\pi/2.
\end{eqnarray}
using the eq.(16), we find,
\begin{equation}
Q = \frac{1}{\pi}(\pi + \theta -sin\theta~ cos\theta).
\end{equation}
Since $g = \Delta Q$ as demonstrated with the Zahed model explicitly ,
we obtain the Haldane statistics parameter in the 3-dimensional chiral
bag case; indeed this is the first time an  example of nontrivial
statistics in dimensions higher than 2 has been discussed. In the end
we also point out another example of Haldane fractional statistics in
higher dimensions, namely the Kondo system.  We note that in the 3+1
dimensional example  we did not make use of the effective single
particle spectrum but obtained $g$ by directly relating it to the
excess charge.  While it is difficult to obtain analytically the
spectrum of states in 3+1 dimensional case, numerical
calculations\cite{kahana} indeed indicate  a flow similar to that in
the Zahed model further buttressing the connection between fractional
charge and the statistics parameter.

To summarize, we have shown that in chiral bag models, the dimension
of the single particle Hilbert space of quarks changes by a fraction,
equal to the fractional charge residing in the bag, when one quark is
added to the system.  This indicates that they satisfy the generalized
Pauli principle as defined by Haldane with the exclusion statistics
parameter $g$.  However there is one important distinction to be
made before accepting this interpretation: Haldane had defined the
generalised Pauli principle for systems of fixed size and boundary
conditions.  In our example, the boundary conditons on the quarks
apparently change and indeed this is the cause of charge
fractionalization.  We should, however, remember, that the boundary
conditions are not being changed by hand but by the dynamics of the
model. The physical condition at the boundary, that is  the
conservation of the axial current, is not changed when a particle is
added. Therefore the phenomenon of charge fractionalization in chiral
bags, which we have shown is the same as the phenomenon of fractional
Pauli blocking, is not put in by hand but is a consequence of the
dynamics of the system.  While the results appear very general, it
should be remembered that the statistics of many body systems that we
are investigating is really vacuum plus one valence particle.   We are
not able, even in the simple 1+1 model, to deal with more than one
kink solution.

Our analysis clearly shows that in chiral bag models, the Haldane
statistical parameter is a topological quantity. The fractional charge
$Q$ of the quark plus the bag and the fractional charge of the soliton
are not independent of each other since the sum has to be equal to the
total integral baryon number which is a global quantum number of the
system and hence conserved.  The solitonic fractional charge is
determined from the topological current and hence $Q$ is a topological
quantity. Clearly then the Haldane statistics parameter $g$ has
topological significance. In a more general context, the same result
may be gleaned from the work of Comtet and Ouvry\cite{comtet} who
relate the second virial coefficient of an anyon gas to the chiral
anomaly in a 1+1 dimensional chiral invariant theory.  Such anomalies
usually arise from topological interactions in the Lagrangian.  The
second virial coefficient of the anyon gas is also related to the
Haldane statistics parameter $g$ albeit in a nontrivial way\cite{ms}.
In the 3+1 dimensional chiral bag model, discussed  above,  the
following features are responsible for the fractional Haldane
statistics: (1) the particles have internal degrees of freedom
(isospin) which couple to the condensate field that can carry a
topological charge and (2) the system is defined in a region with a
boundary and the particles are coupled to the condensates only at the
boundary.

Infact, it is possible that any system in which the above two features
occur the system obeys Haldane fractional statistics.  An interesting
analogous system is the Kondo system.  This system consists of $k$
species of band electrons in three dimensions coupled to a spin $S$
impurity at the origin.  In a recent work\cite{aff} it was shown that
for the low energy physics of this system, the interaction can be
replaced by a dynamical boundary condition.  Thus the problem has
features (1) and (2) if we assume that the electron system is defined
in all space except the origin.  The Kondo problem maps on to a one
dimensional effective system since only the s-wave orbitals see the
impurity. The solution for the low energy physics is a conformal field
theory.  It has been shown that the scaling dimensions of vertex
operators in conformal field theories are related to the Haldane
statistics parameter of the quasiparticles created by
them\cite{ms,kaw}.  In the
overscreened case, ($k>2S$), the solution to the Kondo problem has
operators with scaling dimensions leading to quasiparticles with
fractional exclusion statistics. For example, for $k=2, S=1/2$ there
are spinon excitations with $g=1/2$.  It thus appears that (1) and (2)
are  general features of higher dimensional systems with fractional
statistics.

One of us (R.K.B) thanks The Institute of Mathematical Sciences for
hospitality and the United Nations Tokten programme for financial
assistance.


\bigskip

\begin{references}

\bibitem{haldane}
F.D.M.Haldane, Phys.Rev.Lett. {\bf 67}, 937(1991).

\bibitem{ouvry}
A.Dasni\'{e}res de Veigy and S.Ouvry, Phys.Rev.Lett.{\bf 72},600(1994).

\bibitem{canjohn}
M.D.Johnson and G.S.Canright, Phys.Rev.{\bf B49}, 2947(1994).

\bibitem{ms}
M.V.N.Murthy and R.Shankar, Phys.Rev.Lett.{\bf 72}, 3629(1994).

\bibitem{wu}
Yong-Shi Wu,  Phys.Rev.Lett. {\bf 73},922 (1994).

\bibitem{zahed}
I.Zahed, Phys.Rev. {\bf D30},2647(1984).

\bibitem{comtet}
A.Comtet and S.Ouvry,  Phys.Lett. {\bf 225B},272(1989).

\bibitem{csm}
F.Calogero, J.Math.Phys. {\bf 10}, 2191(1969), ibid {\bf 10},
2197(1969); B.Sutherland, J.Math.Phys. {\bf 12}, 246(1971); ibid {\bf
12}, 251(1971); Phys.Rev. {\bf A4}, 2019(1971); ibid {\bf A5}, 1372
(1972).

\bibitem{ms1}
M.V.N.Murthy and R.Shankar, IMSc Preprint 94/24 (cond-mat/9404096).

\bibitem{wu1}
G.Bernard and Yong Shi Wu, University of Utah Preprint UU-HEP/94-03
(cond-mat/9404015).

\bibitem{vacch}
R.K.Bhaduri, Models of the Nucleon- from quarks to soliton
(Addison-Wesley, 1988)p.386.

\bibitem{cbag}
G.E.Brown and M.Rho, Phys.Lett. {\bf 82B}, 177(1979), G.E.Brown, M.Rho
and V.Vento, Phys.Lett.{\bf 84B}, 383(1979).

\bibitem{jaffe}
J.R.Goldstone and R.Jaffe, Phys.Rev.Lett. {\bf 51}, 1518
(1983),
G.E.Brown, A.S.Goldbaher and M.Rho, Phys.Rev.Lett. {\bf 51}, 747(1983).

\bibitem{kahana}
D.E.Kahana and J.Milana, Nucl.Phys. {\bf A468}, 493(1987).

\bibitem{aff}
I.Afflek and A.W.W.Ludwig, Nucl.Phys. {\bf B360}, 641(1991)

\bibitem{kaw}
T.Fukui and N.Kawakami, Preprint YITP/K-1077(cond-mat/9408015).
\end{references}
\end{document}